\documentclass{article}

\usepackage{epsf,multicol,ifthen}
\usepackage[cp1251]{inputenc}
\usepackage[english,russian]{babel}
\usepackage[dvips]{graphicx}

\renewcommand{\abstractname}{}
\renewcommand{\figurename}{Fig.}
\renewcommand{\refname}{Reference:}
\title{PHENOMENA OF TIME RESONANCES EXPLOSIONS FOR THE COMPOUND-CLOT DECAYS IN HIGH-ENERGY NUCLEAR REACTIONS}

\author{V.S.Olkhovsky\thanks{E-mail: olkhovsk@kinr.kiev.ua}, M.E.Dolinska, S.A.Omelchenko \\
\small\emph{Institute for Nuclear Research, National Academy of Sciences of Ukraine} \\
\small\emph{prosp. Nauki, 47, Kiev-28, 03680, Ukraine}}

\date{}

\begin{document}

\maketitle

\renewcommand{\abstractname}{}
\begin{abstract}
The phenomenon of time resonances (or explosions) can explain the exponential reduction of the energy, which is accompanied for the certain degree by slight fluctuations under some conditions in the range of the energy strongly overlapped compound-resonances. These resonant explosions correspond to formation of several highly-exited non-exponentially decaying nuclear clots (partial compound nuclei consisting of several small groups of projectile nucleons and targets). This paper is a continuation and expansion of theoretical authors' work, which is a more general self-consistent version of the time-evolution approach in comparison with the traditional Izumo-Araseki time compound-nucleus model.
\end{abstract}

\textbf{PACS numbers:} 25.00.00; 25.90.+k; 03.65

\section{INTRODUCTION. PRELIMINARY REMARKS.
\label{sec.introduction}}

There had been observed structureless, exponentially decaying inclusive and not inclusive energy spectra, often accompanied by slight oscillations, throughout the studied range of projectile energies and projectiles, targets and registered final fragments for not very heavy projectiles  (from \emph{p} to $^{20} Ne$) with projectile energies above 0.1 - 1 Gev/nucl (for example, [1-8]). These phenomena for heavier projectiles were observed even for smaller energies (for instance, [9]). In certain degree the fireball models [1,5] can be used successful in the analysis of heavy-ion reactions with energies up to 1 GeV/nucl, and also intranuclear cascade model [10] and in the nuclear fluid model [11] had worked rather well at higher energies, all of them suggesting formation of high-density collision complexes. The difficulties of the fireball models is why a statistical equilibrium is formed even in high excitations (above 100 MeV-nucl). The given paper is continuation and expansion of our previous theoretical paper[12]. Our work is based on the theoretical methods given in [13-17].In this paper is given much more calculation results than in [12], which were carried out on the basis of the author's approach.


\section{JUXTAPOSITION OF ENERGY AND TIME RESONANCES.
\label{sec.2}}
\textbf{2.1. ENERGY RESONANCES.}

Firstly we recall how in the cross section of a reaction $\alpha \rightarrow \beta$,a typical Lorentzian (Breit-Wigner) resonance  is connected with an exponential law of the decay function of the correspondent compound nucleus. The reaction amplitude  $f_{\alpha \beta}(E)$ is represented as

\begin{equation}
f_{\alpha \beta}(E) =
    {\displaystyle\frac{C_{\alpha \beta}}{E-E_{\gamma}+i\Gamma/2} }
\label{eq.1}
\end{equation}
where $C_{\alpha \beta}$ is an almost constant or a smooth function of the final-particle kinetic energy \emph{E} in the region $(E_{\gamma} - \Gamma /2, E_{\gamma} + \Gamma /2)$; in general, it depends on the angular coordinates of the emission direction, $E_{\gamma}$ and $\Gamma$ being the resonance energy and width respectively, let us consider the evolution of the final-particle wave packet.

In the one-dimensional radial asymptotic limit, such form is fair

\begin{equation}
  \Psi_{\beta}(r_{\beta},t)=r_{\beta}^{-1}\int\limits_{0}^{\infty}dE\  g(E)f_{\alpha \beta}(E)\exp[ikr_{\beta}-iEt/\hbar]
\label{eq.2}
\end{equation}
where $g(E)$ is a smooth weight amplitude with an energy spread $\Delta E$ (usually $\Delta E \ll E_{\gamma}$), $m_{\beta}$ and $\gamma_{\beta}$ are the final-particle mass and radial coordinate respectively, and $k = (2m_{\beta} \cdot E)^{1/2} / \hbar$ . The one-dimensional wave packet for short-ranged interactions (including also screened Coulomb potentials), as it was shown in [17], is

\begin{equation}
  \Psi_{\beta}(z_{\beta},t)=\int\limits_{0}^{\infty}dE\  g(E)T_{\alpha \beta}(E)\exp[ikz_{\beta}-iEt/\hbar]
\label{eq.3}
\end{equation}
with $T_{\alpha \beta}(E) = N_{\beta}(E) f_{\alpha \beta}(E)$, $T_{\alpha \beta}(E)$ is the \emph{T}-matrix elements connected with the \emph{S}-matrix elements by known relation $T_{\alpha \beta} = \delta_{\alpha \beta} - S_{\alpha \beta}$ ,  $N_{\alpha \beta}(E)$ is an unessential smooth function of \emph{E}, $z_{\beta}$ is the axis along the direction of the final-particle emission imposed by the registration geometry, $z_{\beta} \geq R_{\beta}$ is the interaction radius in the final channel. The evolution of the particle $\beta$ passing through position $z_{\beta}$ during the unitary time interval, centered at \emph{t} , is described by the probability flux density

\begin{equation}
j_{\beta}(z_{\beta},t) = Re[\Psi_{\beta}(z_{\beta},t)(i\hbar/2m_{\beta})\partial
\Psi_{\beta}^{*}(z_{\beta},t)/\partial z_{\beta}] .
\label{eq.4}
\end{equation}
with the adequate normalization
\begin{equation}
\int\limits_{- \infty}^{\infty}j_{\beta}(z_{\beta},t)dt=1
\label{eq.5}
\end{equation}
In the simplest case can be fix  $z_{\beta} = R_{\beta}$ and use
\begin{equation}
\Psi_{\beta}(R_{\beta},t)=\int\limits_{0}^{\infty}dE\  g(E)T^{\sim}_{\alpha \beta}(E)\exp[-iEt/\hbar]
\label{eq.6}
\end{equation}
where  $T^{\sim}_{\alpha \beta}(E) = T_{\alpha \beta} \exp(ikR_{\beta})$ is a smooth function of \emph{E}: in accordance with the analytical \emph{S}-matrix theory (see [19, 20]), $T_{\alpha \beta}(E)$ contains the factor $\exp(-ikR_{\beta})$ and consequently this factor is being cancelled by $\exp(ikR_{\beta})$ in  $T^{\sim}_{\alpha \beta}(E)$. For condition
\begin{equation}
  \Gamma \ll \Delta E \ll E_{\gamma}
\label{eq.7}
\end{equation}
one can rewrite (6) in the following simplified form
\begin{equation}
 \Psi_{\beta}(R_{\beta},t)=A\int\limits_{0}^{\infty}dE{\displaystyle\frac{\exp [-iEt/\hbar]}{E-E_{\gamma}+i\Gamma/2} }
\label{eq.8}
\end{equation}
where \emph{A} is a constant. In the approximation (7) one can obtain
\begin{equation}
  \Psi_{\beta}(R_{\beta}, t) =
  \left \{
  \begin{array}{lcl}
    B \exp{\displaystyle\frac{-iEt}{\hbar} - (\Gamma/2\hbar)t } &  \mbox{\rm for } & t>0; \\
    0, & \mbox{\rm for } & t<0.
  \end{array}
  \right.
\label{eq.9}
\end{equation}
(moving the lower integration limit in (6) from 0 to $-\infty$ and utilizing the residue theorem). Here \emph{B} is a constant and more precisely there must be $t - t_{in}$   (with  $t_{in} = \hbar (\partial \arg (g)/\partial E)$) instead of \emph{t}.

The emission probability (per a time unit) in the vicinity of the compound nucleus (near $z_{\beta} = R_{\beta}$ )
\begin{equation}
  I (t) =
    \displaystyle\frac{j_{\beta}(R_{\beta}, t)}
    {\displaystyle\int\limits_{-\infty}^{+\infty} j_{\beta}(R_{\beta}, t) \; dt}
\label{eq.10}
\end{equation}
is equal to
\begin{equation}
  I (t) =(\Gamma / \hbar)\exp(- \Gamma t / \hbar)
\label{eq.11}
\end{equation}
In obtaining (11) we took into account that
\begin{equation}
  \lim\limits_{z_{\beta} \to R_{\beta}}(-i \hbar / m_{\beta})T_{\alpha\beta}
    \displaystyle\frac{\partial [\exp(ikz_{\beta})]}
    {\partial z_{\beta}}=\nu T^{\sim}_{\alpha\beta}
\label{eq.12}
\end{equation}

$(\nu = \hbar k/m_{\beta})$. If $\Psi_{\beta} (R_{\beta},t)$ has a form (9), the Fourier-transform of  $\Psi_{\beta}$  is equal to:
\begin{equation}
 \int\limits_{0}^{\infty}dt\ \Psi_{\beta}(R_{\beta},t)\exp [-iEt/\hbar]=B\int\limits_{0}^{\infty}dt\exp[-i(E-E_{\gamma})t/\hbar-(\Gamma/2\hbar)t]=
 \displaystyle\frac{iB}{E-E_{\gamma}+i\Gamma/2}
\label{eq.13}
\end{equation}
which is proportional to the amplitude from (1).

For $z_{\beta} > R_{\beta}$  one can rewrite (3) in a following way:
\begin{equation}
\Psi_{\beta}(z_{\beta},t)=\int\limits_{0}^{\infty}dk\displaystyle\frac{G(k)D(k)}
{(k-k_{0})(k+k_{0})}\exp[ikz_{\beta}-iEt/\hbar]
\label{eq.14}
\end{equation}
with
\[
  G(k)=g(\hbar^{2}k^{2}/2m_{\beta})\displaystyle\frac{dE}{dk},
\]
\[
  D(k)=(2m_{\beta}/\hbar^{2})N_{\beta}(E)C_{\alpha\beta},
\]
\[
  k_{0}=(1/\sqrt{2})\cdot(\sqrt{\sqrt{k^{4}_{\gamma}+\gamma^{2}}+k^{2}_{\gamma}})
  -i(\sqrt{\sqrt{k^{4}_{\gamma}+\gamma^{2}}-k^{2}_{\gamma}}),
\]
\[
  k_{\gamma}=\sqrt{2m_{\beta}E_{\gamma}}/\hbar,\ \gamma=\Gamma m_{\beta}/\hbar^{2}.
\]
Since \emph{G(k}) and \emph{D(k)} are smooth functions of \emph{k} :
\begin{equation}
\Psi_{\beta}(z_{\beta},t)\cong\int\limits_{-\infty}^{\infty}dk\displaystyle\frac{G(k)D(k)}
{(k-k_{0})(k+k_{0})}\exp[ikz_{\beta}-iEt/\hbar]
\label{eq.15}
\end{equation}
under the condition (7) and then, introducing a new variable
\begin{equation}
y=\sqrt{\displaystyle\frac{i\hbar t}{2m_{\beta}}}(k-\displaystyle\frac{m_{\beta}z_{\beta}}{\hbar t})
\label{eq.16}
\end{equation}
and after performing the transformations quite similar to those which were used in [21-23] it is possible to obtain:
\begin{equation}
\Psi_{\beta}(R_{\beta}, t) =
  \left \{
  \begin{array}{lcl}
     0,&  \mbox{\rm for } & z_{\beta} > \nu_{\gamma}t; \\
     const\exp[ik_{\gamma}z_{\beta} - iE_{\gamma}t/\hbar-(\Gamma/2\hbar)(t-z_{\beta}/\nu_{\gamma}) ], & \mbox{\rm for } & z_{\beta} \leq \nu_{\gamma}t.
  \end{array}
  \right.
\label{eq.17}
\end{equation}
with $\nu_{\gamma} = \hbar k/m_{\beta}$. The wave function (17) can be applied for macroscopic distances $z_{\beta}$, near a detector which registers particles $\beta$.

Let us remember that an exponential law (11) and also the asymptotic (17) are valid only under conditions (7), i.e. when all energies (or continuum states) around $E_{\gamma}$ are completely populated in the large region with the width $\Delta E \gg \Gamma$.  If, on the contrary,
\begin{equation}
\Delta E \ll \Gamma
\label{eq.18}
\end{equation}
the emission probability is non-exponential and does essentially depend on $\Delta E$ and the form of \emph{g(E)}. If one will take the Lorentzian form also for \emph{g(E)}, i.e
\begin{equation}
g(E^{\prime})= \displaystyle\frac{g_{0}}{E^{\prime}-E+i\Delta E /2}
\label{eq.19}
\end{equation}
with $g_{0} = const$ or smooth inside $\Delta E$ under conditions (7) and (18), instead of (17) the expression
\begin{equation}
\Psi_{\beta}(R_{\beta}, t) =
  \left \{
  \begin{array}{lcl}
     0,&  \mbox{\rm for } & z_{\beta} > \nu_{\gamma}(t-t^{0}_{in}); \\
     \displaystyle\frac{const}{E-E_{\gamma}+\displaystyle\frac{i\Gamma}{2}}
     \exp[ikz_{\beta} - \displaystyle\frac{iE_{\gamma}t}{\hbar} -\displaystyle\frac{\Delta E}{2\hbar}
     (t-\displaystyle\frac{z_{\beta}}{\nu}-t^{0}_{in}) ], & \mbox{\rm for } & z_{\beta} \leq \nu_{\gamma}(t-t^{0}_{in}).
  \end{array}
  \right.
\label{eq.20}
\end{equation}
(with $\nu = \hbar k/m_{\beta}$ and  $t_{in}^{0}=\hbar \displaystyle\frac{\partial \arg g_{0}}{\partial E}$) will be obtained. The cross section  $\sigma_{\alpha\beta}$ , which is proportional to
\begin{equation}
\sigma_{\alpha\beta}\sim \int\limits_{t_{min}}^{\infty}dt j_{\beta}(z_{\beta},t)
\label{eq.21}
\end{equation}
where $t_{min} = z_{\beta} / \nu + t_{in}^{0}$, ($t_{min}, \infty$) being the operative registration time interval of detector, after integrating in (21) acquires the Breit-Wigner form
\begin{equation}
\sigma_{\alpha\beta}=|f_{\alpha\beta}|^{2}=\displaystyle\frac{const}{(E-E_{\gamma})^{2}+\Gamma^{2}/4}.
\label{eq.22}
\end{equation}

\textbf{2.2. TIME RESONANCES.}

Let us use the same method in the case when
\begin{equation}
T^{\sim}_{\alpha \beta}(E) = T^{\sim n}_{\alpha\beta}\exp(-E\tau_{n}/2\hbar +iEt_{n}/\hbar)
\label{eq.23}
\end{equation}
with $E_{min} <E< \infty $, where $\tau_{n}$ and $t_{n}$ are constants (with the time dimension), $\tau_{n}$ and $t_{n}$ determine the  exponential dependence on energy for the correspondent cross section and the linear dependence on energy for the amplitude phase respectively. $T^{\sim n}_{\alpha \beta}$ is a constant or a very smooth function (inside $\Delta E$) of the final-particle energy \emph{E}. Here the fine resonance structure of  $T^{\sim n}_{\alpha \beta}$ is not taken explicitly into account yet, supposing it totally smoothed out by the energy spread (and resolution) $\Delta E$, and also assuming that $\Delta E<< 2\hbar/\tau_{n}$ .

In this case, one can write instead of (6) and (8-9)
\begin{equation}
 \Psi_{\beta}(R_{\beta},t)\approx \int\limits_{E_{min}}^{\infty}dE^{\prime}A^{\prime}
 \exp[-E^{\prime}\tau_{n}/2\hbar + iE^{\prime}(t_{n}-t)/\hbar]
\label{eq.24}
\end{equation}
with $A^{\prime}=T^{\sim n}_{\alpha \beta}g(E^{\prime})$. Using the simplest rectangular form of $g(E^{\prime})$, i.e.
\begin{equation}
  g(E^{\prime}) =
  \left \{
  \begin{array}{lcl}
     (\Delta E)^{-1/2}\exp{(i \arg g)}, &  \mbox{\rm for } & E_{min}\leq E-(\Delta E)/2 < E^{\prime}< E+(\Delta E)/2; \\
    0, & \mbox{\rm for } & E^{\prime} < E-(\Delta E)/2 \;\; and \;\; E^{\prime}> E+(\Delta E)/2.
  \end{array}
  \right.
\label{eq.25}
\end{equation}
with $\arg g$ being a smooth function of \emph{E} inside $\Delta E$, one gets
\begin{equation}
\;
  \begin{array}{lcl}
     \Psi_{\beta}(R_{\beta},t)=\displaystyle\frac{const}{t-t_{n}+i\tau_{n}/2}
     \exp[E(- \tau_{n}/2 +i(t_{n}-t))/\hbar]\cdot\\
    \cdot[\exp[\Delta E(- \tau_{n}/2 +i(t_{n}-t)/2\hbar)]-\exp[-\Delta E(- \tau_{n}/2 +i(t_{n}-t)/2\hbar)]]
  \end{array}
\label{eq.26}
\end{equation}
When all energies levels in the large area beginning with $E_{min}$, are completely filled, i.e.
\begin{equation}
  \;
  \left \{
  \begin{array}{lcl}
     (E + \Delta E/2)\tau_{n}/2\;\;\rightarrow\; \infty &  \mbox{\rm and }  \\
    \;\; E + \Delta E/2\;\;\rightarrow\; E_{min},
  \end{array}
  \right.
\label{eq.27}
\end{equation}
one arrives to
\begin{equation}
\Psi_{\beta}(R_{\beta},t)=\displaystyle\frac{const}{t-t_{n}+i\tau_{n}/2}
\exp[E_{min}(- \tau_{n}/2 +i(t_{n}-t))/\hbar].
\label{eq.28}
\end{equation}
It is natural such a behavior of $\Psi_{\beta}(R_{\beta},t)$ to call a \emph{time resonance} because of the Lorentzian form of the factor $\displaystyle\frac{1}{t-t_{n}+i\tau_{n}/2}$ in (28), \emph{or an explosion} (for small values of $\tau_{n}$). And inversely, if $\Psi_{\beta}(R_{\beta},t)$ has a form (28), the Fourier-transform of $\Psi_{\beta}(R_{\beta},t)$ will be equal to
\begin{equation}
\;
  \begin{array}{lcl}
     \int\limits_{-\infty}^{\infty}dt \Psi_{\beta}(R_{\beta},t)
     \exp[iEt/\hbar]=\\
    const\cdot\exp[-E \tau_{n}/2\hbar +iEt_{n}/\hbar+E_{min} \tau_{n}/2\hbar]
  \end{array}
\label{eq.29}
\end{equation}
which is proportional to the amplitude (23).

For $z_{\beta} > R_{\beta}$  it is possible to rewrite (3) in a following way:
\begin{equation}
  \;
  \begin{array}{lcl}
     \Psi_{\beta}(z_{\beta},t)=\int\limits_{E_{min}}^{\infty}dE^{\prime}
     f^{n}_{\alpha\beta}N_{\beta}\exp(ikz_{\beta})g(E^{\prime})\cdot\\
    \cdot\exp[-E^{\prime}\tau_{n}/2\hbar +iE^{\prime}(t_{n}-t)/\hbar].
  \end{array}
\label{eq.30}
\end{equation}
If the narrow energy spread $(\Delta E << E)$, using the function (19) for $g(E^{\prime})$ and introducing a new variable
\begin{equation}
y^{\prime}=\sqrt{\displaystyle\frac{i\hbar(t-t_{n}-i\tau_{n}/2)}{2m_{\beta}}}
[k-\displaystyle\frac{m_{\beta}z_{\beta}}{\hbar (t-t_{n}-i\tau_{n}/2)}],
\label{eq.31}
\end{equation}
instead of (16), one finally obtains:
\begin{equation}
\Psi_{\beta}(R_{\beta}, t) =
  \left \{
  \begin{array}{lcl}
     0,&  \mbox{\rm for } & z_{\beta} > \nu(t-t_{n}-t^{0}_{in}); \\
     const\cdot
     \exp[ikr - \displaystyle\frac{iE(t-t_{n}-t^{0}_{n}-i\tau_{n}/2)}{\hbar}- \Delta EA(t)], & \mbox{\rm for } & z_{\beta} \leq \nu(t-t_{n}-t^{0}_{in}).
  \end{array}
  \right.
\label{eq.32}
\end{equation}
with
\[
  A(t)=[t-t_{n}-t^{0}_{n}-z_{\beta}/\nu-i\tau_{n}/2]/2\hbar
\]
after the transformations similar to those which were made for obtaining (17) and (20). The cross section, defined by the formula (21), will have an exponential form:
\begin{equation}
\sigma_{\alpha\beta}=|f_{\alpha\beta}|^{2}=const\cdot\exp(-E\tau_{n}/\hbar).
\label{eq.33}
\end{equation}
When $T^{\sim}_{\alpha\beta}$  or $f_{\alpha\beta}$ has a more general form like
\begin{equation}
f_{\alpha\beta}=\sum\limits_{n=1}^{\nu}f^{n}_{\alpha\beta}
\exp[-E\tau_{n}/2\hbar+iEt_{n}/\hbar].
\label{eq.34}
\end{equation}
with several terms $(\nu = 2,3,…)$, the cross section $\sigma_{\alpha\beta}=|f_{\alpha\beta}|^{2}$ contains not only exponentially decreasing terms but also oscillating terms with factors $\cos[E(t-t_{n^{\prime}})/\hbar]$ and $\sin[E(t-t_{n^{\prime}})/\hbar]$.)/ ]. In the case of two terms $(\nu =2)$ in (34) formula  (33) pass to a kind
\begin{equation}
\;
  \begin{array}{lcl}
     \sigma_{\alpha\beta}=|f^{1}_{\alpha\beta}|^{2}\exp(-E\tau_{1}/\hbar)+
     |f^{2}_{\alpha\beta}|^{2}\exp(-E\tau_{2}/\hbar)\\
    +2Re\{f^{1}_{\alpha\beta}f^{2}_{\alpha\beta}\cdot\exp[iE(t_{1}-t_{2})/\hbar
    -E(\tau_{1}+\tau_{2})/2\hbar]\}.
  \end{array}
\label{eq.35}
\end{equation}
(where the terms with $\Delta E$ are neglected if the conditions $\Delta Et_{n}<<E\tau_{n}$  and  $\Delta E\tau_{n}<<Et_{n}$  are supposed).

\section{COMPOUND NUCLEI PROPERTIES WHICH CORRESPOND AND DO NOT CORRESPOND TO TIME RESONANCES.
\label{sec.3}}
\textbf{3.1. COMPOUND-NUCLEUS DECAY EVOLUTION IN THE REGION OF A TIME RESONANCE.}

The compound-nucleus surviving evolution (at instant \emph{t} during the life and decay after the formation) can be described by the following function:
\begin{equation}
L^{c}(t) = 1 - \int\limits_{t_{0}}^{t}dt I(t)
\label{eq.36}
\end{equation}
where \emph{I(t)} is defined by (10). The initial point $t_{0}$ of time flowing seems naturally to be chosen at moment $t_{in}^{0}$, thus supposing $t_{in}^{0}=0$. However, the uncertainly $\delta t =  \hbar/\Delta E$ of the initial wave packet duration before the collision must be taken into account. So $t_{0}\cong t^{0}_{n}-\delta t= -\delta t=-\hbar/\Delta E.$

In the region of an isolated energy resonance the function \emph{I(t)} is described by exponential formula (11) and the function $L^{c}(t)$ is also an exponential one:
$L^{c}(t) = 1 - \int\limits_{t_{0}}^{t}dt(\Gamma/\hbar)\exp(-\Gamma t/\hbar)=\exp(-\Gamma t/\hbar).$

\begin{figure}[htbp]
\centerline{
\includegraphics[width=55mm]{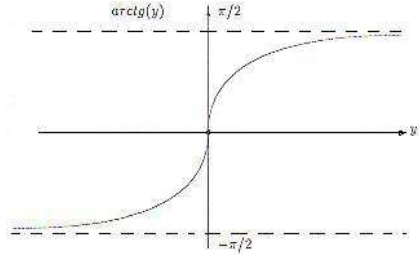}}
\caption{\small
The function $\arctan(y)$ for $2t_{0}/\tau_{n}\rightarrow -\infty$
\label{fig.example.1}}
\end{figure}

\begin{figure}[htbp]
\centerline{
\includegraphics[width=75mm]{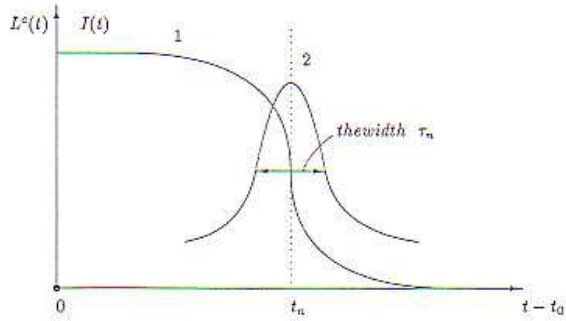}}
\caption{\small
$L^{c}(t)$ (curve 1) and $I(t)$ (curve 2).
\label{fig.example.2}}
\end{figure}

In the region of a time resonance (28) the function $L^{c}(t)$ is essentially non-exponential even at the approximation $t_{0} = 0$. The qualitative form of $L^{c}(t)$ can be illustrated with the help of the very simplified examples, using (28) at a very narrow interval near $t = t_{n}$ , and also for all values of \emph{t}, when
\begin{equation}
j_{\beta}(R_{\beta},t)= Re[\Psi_{\beta}(R_{\beta},t)\cdot (i\hbar/m_{\beta})\cdot\lim\limits_{z_{\beta}\rightarrow R_{\beta}}
\delta\Psi^{\star}_{\beta}/\partial(z_{\beta},t)/\partial(z_{\beta})]
\cong\overline{\nu}|\Psi_{\beta}(R_{\beta},t)|^{2}
\label{eq.37}
\end{equation}
where $\overline{\nu}$  is defined by the integral theorem about the average value, namely by the relation
\begin{equation}
\int\limits_{E_{min}}^{\infty}dE\nu A\exp(-E\tau_{n}/2\hbar)=\overline{\nu}
\int\limits_{E_{min}}^{\infty}dE A\exp(-E\tau_{n}/2\hbar)
\label{eq.38}
\end{equation}
($\nu$ here appeared after applying (12)). Then
\begin{equation}
  \;
  \begin{array}{lcl}
     I(t)=\displaystyle\frac{j_{\beta}(R_{\beta},t)}
     {\int\limits_{-\infty}^{+\infty}dtj_{\beta}(R_{\beta},t)}
     \cong\displaystyle\frac{[(t-t_{n})^{2}+\tau_{n}^{2}/4]^{-1}}
     {\int\limits_{-\infty}^{+\infty}dt[(t-t_{n})^{2}+\tau_{n}^{2}/4]^{-1}}=\\
    =(\tau_{n}/2\pi)\displaystyle\frac{1}{(t-t_{n})^{2}+\tau_{n}^{2}/4}
  \end{array}
\label{eq.39}
\end{equation}
and
\begin{equation}
  L^{c}(t)=1-(1/\pi)[\arctan(y)]
  \begin{array}{lcl}
     y=2(t-t_{n}-t_{0})\tau_{n}\\
    y=2(2t_{0})\tau_{n}
  \end{array}
\label{eq.40}
\end{equation}
Since the curve $arctan(y)$ has the form which is illustrated by Fig.~\ref{fig.example.1}, in the case of $2t_{0}/\tau_{n}\rightarrow -\infty$ ($\tau_{n}$ are small) the function $L^{c}(t)$ has a form, depicted in Fig.~\ref{fig.example.2} (curve 1).
%

\renewcommand{\figurename}{Fig.}

In this case
\[
  L^{c}(t)=1- \pi[\arctan(2(t-t_{n}-t_{0})\tau_{n})-\pi/2]\; \ \ \ \ \ \ \ \ \ \ \  \ \ \ \ \ \ \ \ \ \ \ \ (40a)
\]
\[
L^{c}(t) =
  \left \{
  \begin{array}{lcl}
     1,&  \mbox{\rm when } & t=0 \ \ \ (with \ t_{n}+\hbar/\Delta E \rightarrow \infty)\ \ \ and \\
     0, & \mbox{\rm when} & t\rightarrow \infty \ \ \ \ \ \ \ \ \ \ \ \
      \ \ \ \ \ \ \ \ \ \ \ \  \ \ \ \ \ \ \ \ \ \ \ \  \ \ \ \ \ \ \ \ \ \ \ \ (40b)
  \end{array}
  \right.
\]
From the simple form of Fig.2, one can see that $t_{n}$ can be interpreted as the Poincare period of internal motions in the compound nucleus (after its formation and before its decay) when $t_{n} >> \tau_{n}$. Such behavior of $L^{c}(t)$ was investigated in [14,16,18]. If one will take into account the compound-resonance structure of $T_{\alpha\beta}$ explicitly, the strongly non-exponential form of $L^{c}(t)$ and \emph{I(t)}, like depicted in Fig.2, will take place for the case of strongly overlapped energy resonances when
\begin{equation}
\Gamma_{JS\Pi} \ll N_{JS\Pi}/2\pi\rho_{JS\Pi}
\label{eq.41}
\end{equation}
($\Gamma_{JS\Pi}$ and $\rho_{JS\Pi}$ are the mean resonance width and level density, $N_{JS\Pi}$ is the number of open channels, $JS\Pi$ being total momentum, spin and parity quantum numbers respectively). The compound-nucleus decaying small probability for $t < t_{n}$ (inside the Poincare cycle) can be understood as a consequence of the multiple internal transitions between the strongly non-orthogonal wave functions of the meta-stable states in the region of the strongly overlapped energy resonances.

In the case of a several time resonances, it can signify the superposition of several strongly overlapped energy-resonance groups with different values of $JS\Pi$ of the same compound nucleus or the formation of several partial compound nuclei with various numbers of participating nucleons.

\textbf{3.2. A POSSIBILITY OF EXPLAINING A TIME-RESONANCE STRUCTURE OF AMPLITUDES FOR MANY CHANNELS AT THE RANGE OF OVERLAPPED ENERGY RESONANCES.}

How can explain manipulations with the comparatively smooth energy behavior of expressions (23) and (34) for $T_{\alpha\beta}$ or $f_{\alpha\beta}$ which correspond to time resonances and also to experimental data on cross sections, although really the amplitudes are strongly fluctuating in the region of a lot of the overlapped energy resonances with extremely high densities? At first sight , in the region of high energies the energy resonance structure has to vanish not only because of the "smoothing" by energy spreads (since $\Delta E >> \Gamma_{JS\Pi}$, $\rho^{-1}_{JS\Pi}$) but also de facto because of the strong decreasing of the probability of the formation of intermediate long-living multi-nucleon states. The compound-resonance density is very fast increasing with energy, beginning from low-energy well-resolved energy resonances where various versions of the Fermi-gas model with shell-model and collective-model corrections are rather successful. Only near 30-40 MeV/nucleon in the compound-nucleus system one can expect saturation effects and further the strong decreasing. However, just for these energies, resonances of another nature can appear. These resonances are connected with local excitations of long-living intermediate multi-quark-gluon states of baryon subsystems (see [24]).

Let us consider the possibility of above-mentioned explaining the time-resonance structure of amplitudes attentively, limiting ourselves by considering only the partial $JS\Pi$-amplitudes $T^{JS\Pi}_{\alpha\beta} = \delta_{\alpha\beta} - S^{JS\Pi}_{\alpha\beta}$ , $S^{JS\Pi}_{\alpha\beta}$ being S-matrix element.

For sufficiently high energies, if one neglects bound-state, anti-bound-state and threshold singularities, the \emph{S}-matrix can be described by the following analytic expression [14-18, 25]:
\begin{equation}
\widehat{S}=\widehat{U}\widehat{S}_{res}\widehat{U}^{T},\;\; \widehat{S}_{res}=
\prod\limits_{n}(I-\displaystyle\frac{i\Gamma_{n}\widehat{P}_{n}}{\varepsilon -\varepsilon_{n}+i\Gamma_{n}/2})
\label{eq.42}
\end{equation}
where the indexes $JS\Pi$ are omitted for the simplicity, the unitary background (non-resonant) matrix \emph{U} and the resonance projection matrix $\widehat{P_{n}}$ ($\widehat{P_{n}}= \widehat{P_{n}}^{+}= \widehat{P_{n}}^{2}$, Trace $\widehat{P_{n}}= 1$)slowly changing with the total energy $\varepsilon$ or are almost independent on $\varepsilon$ , $\widehat{U}^{T}$ being the matrix transposed to $\widehat{U}$ . Under the simplest Baz'-Newton conditions (see [14-18]) when fluctuations of $\widehat{P_{n}}$  can be neglected $(\widehat{P_{n}}=<\widehat{P}>)$, the \emph{S}-matrix (42) acquires the following form :
\begin{equation}
\widehat{S}=\widehat{S}_{b}-\widehat{a}(1-\prod\limits_{n}\displaystyle\frac
{\varepsilon-\varepsilon_{n}-i\Gamma_{n}/2}{\varepsilon-\varepsilon_{n}+i\Gamma_{n}/2})
\label{eq.43}
\end{equation}
with $\widehat{S}_{b}=\widehat{U} \widehat{U}^{T}$ and $\widehat{a}=\widehat{U}<\widehat{P}>\widehat{U}^{T}$ . The energy-averaged \emph{S}-matrix $<\widehat{S}>_{\Delta \varepsilon}$ is [14-16]

$<\widehat{S}>_{\Delta \varepsilon}=\widehat{S}_{b}-\widehat{a}[1-\exp(-\pi\Gamma/\rho)]$ for unresolved resonances $(\Delta E >> \rho^{-1}, \Gamma)$ and fluctuating (or compound-nucleus) \emph{S}-matrix $\widehat{S}^{c}$ is
\begin{equation}
\widehat{S}^{c}=\widehat{S}-<\widehat{S}>_{\Delta \varepsilon}=\widehat{a}
[\prod\limits_{n}\displaystyle\frac
{\varepsilon-\varepsilon_{n}-i\Gamma_{n}/2}{\varepsilon-\varepsilon_{n}+i\Gamma_{n}/2}
-\exp(-\pi\Gamma\rho)]
\label{eq.44}
\end{equation}

Let us repeat that $\widehat{S}_{b}$ and $\widehat{a}$ are almost independent on energy (slowly changing with energy). For strongly overlapped resonances when $\pi\Gamma\rho >> 1$
\[
  \widehat{S}^{c} \rightarrow \widehat{a}
\prod\limits_{n}\displaystyle\frac
{\varepsilon-\varepsilon_{n}-i\Gamma_{n}/2}{\varepsilon-\varepsilon_{n}+i\Gamma_{n}/2}
\; \ \ \ \ \ \ \ \ \ \ \  \ \ \ \ \ \ \ \ \ \ \ \
\; \ \ \ \ \ \ \ \ \ \ \  \ \ \ \ \ \ \ \ \ \ \ \ (44a)
\]
and the energy-averaged compound-nucleus cross section $<\sigma^{c}_{\alpha\beta}>_{\Delta \varepsilon}$ is evidently proportional to $|a_{\alpha\beta}|^{2}$:
\begin{equation}
<\sigma^{c}_{\alpha\beta}>  \sim  <|S^{c}_{\alpha\beta}|^{2}>_{\Delta \varepsilon}=|a_{\alpha\beta}|^{2}
\label{eq.45}
\end{equation}
(here and below we continue omitting indexes $JS\Pi$). If the initial energy of projectiles is fixed and therefore the total energy $\varepsilon$  is also fixed (with the accuracy $\Delta \varepsilon$) the cross section (45) can be rewritten in the form
\[
 <\sigma^{c}_{\alpha\beta}>_{\Delta E}\ \   \sim \ \  <|S^{c}_{\alpha\beta}|^{2}>_{\Delta \varepsilon}\ \ =\ \ <|a_{\alpha\beta}|^{2}>_{\Delta \varepsilon}\ \  \cong|\ \  a_{\alpha\beta}|^{2}
 \; \ \ \ \ \ \ \ \ \ \ \  \ \ \ \ \ \ \ \ \ \ \ \
\; \ \ \ \ \ \ \ \ \ \ \  \ \ \ \ \ \ \ \ \ \ \ \ (45a)
\]
where $\Delta E$ is defined by $\Delta\varepsilon$ and the energy resolution of the detector of final fragments.

As it is shown in [18, ref.1], see also [14] the correspondent to (45)-(45a) energy-averaged compound-nucleus time delay and variance of the compound-nucleus delay distribution are defined by the following general relations
\begin{equation}
<{\tau^{c}}^{JS\Pi}_{ik}>  =  <|{S^{c}}^{JS\Pi}_{ik}|^{2}>\hbar
\displaystyle\frac{\partial\arg {S^{c}}^{JS\Pi}_{ik}}{\partial E}/<|{S^{c}}^{JS\Pi}_{ik}|^{2}>
\label{eq.46}
\end{equation}
and
\begin{equation}
\;
  \begin{array}{lcl}
     D{\tau^{c}}^{JS\Pi}_{ik}=\displaystyle\frac{\hbar^{2}<(\partial|{S^{c}}^{JS\Pi}_{ik}|/\partial E)^{2}>_{\Delta E}}{<|{S^{c}}^{JS\Pi}_{ik}|^{2}>_{\Delta E}}

    \\
    +\displaystyle\frac{\hbar^{2}<|{S^{c}}^{JS\Pi}_{ik}|^{2}(\partial \arg {S^{c}}^{JS\Pi}_{ik}/\partial E)^{2}>_{\Delta E}}{<|{S^{c}}^{JS\Pi}_{ik}|^{2}>_{\Delta E}}\ \ -\ \ <{\tau^{c}}^{JS\Pi}_{ik}>^{2}
  \end{array}
\label{eq.47}
\end{equation}
respectively (energy \emph{E} is the final-fragment kinetic energy).

Before using (46)-(47) we shell derive the sum rule connects the compound-nucleus mean time delay, the variance of the compound-nucleus time-delay distributions with the main compound-nucleus resonance characteristics.

For the simplicity here and further we omit in all the formulas the indexes $JS\Pi$.

After a series of transformations of Trace $\widehat{R}$ with utilization of the relation
\begin{equation}
    {\displaystyle\frac{d}{dE} }\sum\limits_{k}{S_{ik}S^{*}_{ik}}=0 ,
\label{eq.48}
\end{equation}
following directly from the differentiation of the unitarity relation $\sum\limits_{k}=1$ for the \emph{S}-matrix, one can obtain such relation

\begin{equation}
    Trace \widehat{R}=\hbar^{2}\{\sum\limits_{ik}[({\displaystyle\frac{dT_{ik}}{dE})^{2}+|T_{ik}|^{2}
     (\displaystyle\frac{d\ arg T_{ik}}{dE})^{2}}]+i \sum\limits_{ik}[\displaystyle\frac{d\ arg T_{ik}}{dE}
     \ \displaystyle\frac{d |T_{ik}|^{2}}{dE}\ + |T_{ik}|^{2}\displaystyle\frac{d^{2}\ arg T_{ik}}{dE}]
     +i\sum\limits_{i} \displaystyle\frac{d^{2} Im\ S_{ii}}{dE^{2}}\} ,
\label{eq.49}
\end{equation}
where $T_{ik}=\ \delta_{ik}-S_{ik}$ is now the element of the partial \emph{T}-matrix.

By averaging relation (49) on the energy spread of wave packets in the initial channel, one can write
\begin{equation}
    <Trace\widehat{R}>=\ {\displaystyle\frac{\int|g(E)|^{2}\ Trace\widehat{R}\ dE}{\int|g(E)|^{2}dE} } ,
\label{eq.50}
\end{equation}
where the weight function $|g(E)|^{2}$  is normalized by the condition  $\int|g(E)|^{2}dE\ =\ 1$.

Integrating the term with $\displaystyle\frac{d^{2}\ arg T_{ik}}{dE^{2}}$  by parts in (50), one can easily be convinced that the second sum $\sum\limits_{ik}$ in (50) for smooth weight functions $|g(E)|^{2}$  vanishes and can consequently be ignored. Moreover, the sum $\sum\limits_{i}$ in (50) can also be neglected since it is a quantity of the order of at least $(\Delta E)^{-1}$ as it can be easily seen after integrating (especially for the rectangular form of $|g(E)|^{2}$. Thus, in this approximation one can write
\begin{equation}
    <Trace \widehat{R}>=\ Re<Trace \widehat{R}>=\ \hbar^{2}\sum\limits_{ik}[<(\displaystyle\frac{dT_{ik}}{dE})^{2}>+
    <|T_{ik}|^{2}(\displaystyle\frac{d\ arg T_{ik}}{dE})^{2}>] ,
\label{eq.51}
\end{equation}
As one can see from the expressions (5)-(6) for the collision duration variance $D\tau_{ik}$  (here for the partial collision variance with the omitted index \emph{J}), with sufficiently large collision durations, one can rewrite relation (51) in the form
\begin{equation}
    Re<Trace \widehat{R}>=\ \sum\limits_{ik}[D\tau_{ik}\ +\ <\tau_{ik}>^{2}]\ /\ <|T_{ik}|^{2}> .
\label{eq.52}
\end{equation}
Now we shall study $\widehat{R}$ and $<Trace\widehat{R} >$, using the Simonius representation (42) of the \emph{S}-matrix (as usually here omitting the indexes \emph{J}). Since, evidently, $<Trace\widehat{R} >=<Trace\widehat{R}_{res} >$ with $\widehat{R}_{res}=-\hbar^{2}\widehat{S}_{res}(d^{2}\widehat{S}_{res}/dE^{2})$ and
$\widehat{S}_{res}=\prod\limits_{\nu=1}^{n}(1-\displaystyle\frac{i\Gamma_{\nu} \widehat{P}_{\nu}}{E-E_{\nu}
+i\Gamma_{\nu}/2})$ it is sufficient to study $\widehat{R}_{res}$.

The principally simple but rather bulky calculations give
\begin{equation}
\;
  \begin{array}{lcl}
    Trace \widehat{R}_{res}=\ -2i\hbar^{2}\sum\limits_{m}\displaystyle\frac{\Gamma_{m}\cdot Trace \widetilde{P}_{m}}{[(E-E_{m})^{2}+\Gamma_{m}^{2}/4](E-E_{m}-i\Gamma_{m}/2)}+\\
    +\ 2\hbar^{2}\sum\limits_{m'<m}\displaystyle\frac{\Gamma_{m}\Gamma_{m'}\cdot Trace \widetilde{P}_{m}\cdot\widetilde{P}_{m'}}{[(E-E_{m})^{2}+\Gamma_{m}^{2}/4][(E-E_{m'})^{2}+\Gamma_{m'}^{2}/4]},
 \end{array}
\label{eq.53}
\end{equation}
where $\widetilde{P}_{m}=\ \widehat{s}_{m}\widehat{P}_{m}\widehat{s}_{m}^{*},\ \widehat{s}_{m}=\prod\limits_{\nu=1}^{m-1}(1-\displaystyle\frac{i\Gamma_{\nu} \widehat{P}_{\nu}}{E-E_{\nu}
+i\Gamma_{\nu}/2})$ ,    with $\widehat{s}_{m}\widehat{s}_{m}^{*}=1$  ,  ( $\widehat{P}_{m}$ is also a projector since it is a unitary transform of the projector $\widehat{P}_{m}$ ) and we have used the relations
                 $Trace\widehat{P}_{m}\cdot\widehat{P}_{m'} = Trace\widehat{P}_{m'}\cdot\widehat{P}_{m} ,  \ \ Trace\widehat{P}_{m}\cdot\widehat{P}_{m'}\cdot\widehat{P}_{m} = Trace\widehat{P}_{m'}\cdot\widehat{P}_{m}$ ,
which are the evident consequences of the trace cyclicity and projector properties. In order to analyze $ Trace\widehat{P}_{m}\cdot\widehat{P}_{m'}$   we shall use the representation adopted in [18],
\begin{equation}
    (\widehat{P}_{m})_{\nu\nu'}=x_{m,\nu}\cdot x_{m,\nu}^{*},
\label{eq.54}
\end{equation}
with $\sum\limits_{\nu=1}^{n}|x_{m,\nu}|^{2}=1$ . Then
\begin{equation}
    (\widetilde{P}_{m})_{\nu\nu'}=(\widehat{y}_{m})_{\nu}\cdot (\widehat{y}_{m}^{*})_{\nu '},
\label{eq.55}
\end{equation}
where $(y_{m})_{\nu}=\sum\limits_{\nu '}s_{m,\nu \nu '}\cdot x_{m,\nu '}$ and $\sum\limits_{\nu=1}^{n}|(y_{m})_{\nu}|^{2}=1$.

Therefore $\widetilde{P}_{m}=1$ and $Trace\widetilde{P}_{m}\widetilde{P}_{m'} $     becomes      $Trace\widetilde{P}_{m}\cdot\widetilde{P}_{m'} = |\sum\limits_{\nu}A_{mm',\nu}|^{2}$, where

$A_{mm',\nu}=(\widehat{y}_{m})_{\nu}\cdot(\widehat{y}_{m'}^{*})_{\nu}$. Hence
\begin{equation}
    0<Trace\widetilde{P}_{m}\cdot\widetilde{P}_{m'}=|\sum\limits_{\nu}A_{mm',\nu}|^{2}\leq 1,
\label{eq.56}
\end{equation}

Rewriting (53) in the averaged (on the energy spread of wave packets in the initial channel)  form
\begin{equation}
\;
  \begin{array}{lcl}
    <Trace \widehat{R}_{res}>=Re<Trace \widehat{R}_{res}>=
    \ \{-2i\hbar^{2}\sum\limits_{m}\displaystyle\frac{\Gamma_{m}}{[(E-E_{m})^{2}+\Gamma_{m}^{2}/4](E-E_{m}-i\Gamma_{m}/2)}+\\
    +\ 2\hbar^{2}\sum\limits_{m'<m}\displaystyle\frac{\Gamma_{m}\Gamma_{m'}\cdot Trace \widetilde{P}_{m}\cdot\widetilde{P}_{m'}}{[(E-E_{m})^{2}+\Gamma_{m}^{2}/4][(E-E_{m'})^{2}+\Gamma_{m'}^{2}/4]}\}>,
 \end{array}
\label{eq.57}
\end{equation}
we get
\begin{equation}
\;
  \begin{array}{lcl}
     <\sum\limits_{m}\displaystyle\frac{(-2i\hbar^{2})\Gamma_{m}}{[(E-E_{m})^{2}+\Gamma_{m}^{2}/4](E-E_{m}-i\Gamma_{m}/2)}>\approx\\
    \approx\ \int\limits_{-\infty}^{+\infty}\displaystyle\frac{d\varepsilon}{\overline{D}}
    \displaystyle\frac{(-2i\hbar^{2})\overline{\Gamma}}{[\varepsilon^{2}+\overline{\Gamma}^{2}/4]
    (\varepsilon -i\overline{\Gamma}/2)}=\displaystyle\frac{4\pi\hbar^{2}}{\overline{D\Gamma}},
\end{array}
\label{eq.58}
\end{equation}
\[
   <2\hbar^{2}\sum\limits_{m'<m}\displaystyle\frac{\Gamma_{m}\Gamma_{m'}\cdot Trace \widetilde{P}_{m}(E)\cdot\widetilde{P}_{m'}(E)}
   {[(E-E_{m})^{2}+\Gamma_{m}^{2}/4][(E-E_{m'})^{2}+\Gamma_{m'}^{2}/4]}>\approx
\]
\[
  \approx 2\hbar^{2}\displaystyle\frac{\overline{\Gamma}^{2}}{\overline{D}^{2}}
  \int\limits_{-\infty}^{+\infty}d\varepsilon\int\limits_{-\infty}^{\varepsilon}d\varepsilon'
  \displaystyle\frac{(Trace \widetilde{P}(\varepsilon,E)\cdot\widetilde{P}(\varepsilon',E))}
  {[\varepsilon^{2}+\overline{\Gamma}^{2}/4][\varepsilon'^{2}+\overline{\Gamma}^{2}/4]}=
\]
\[
  = 2i\hbar^{2}\displaystyle\frac{\overline{\Gamma}^{2}}{\overline{D}^{2}}
  \int\limits_{-\infty}^{+\infty}d\varepsilon
  \displaystyle\frac{[ln\displaystyle\frac{\varepsilon+i\overline{\Gamma}/2}
  {\varepsilon-i\overline{\Gamma}/2}](Trace \widetilde{P}(\varepsilon,E)\widetilde{P}(\varepsilon'_{c},E))}
  {[\varepsilon^{2}+\overline{\Gamma}^{2}/4]}=
\]
\begin{equation}
\;
  \begin{array}{lcl}
     =\displaystyle\frac{2i\hbar^{2}}{\overline{D}}<(Trace \widetilde{P}(\varepsilon_{c},E)\widetilde{P}(\varepsilon'_{c},E))>
     \int\limits_{-\infty}^{+\infty}d\varepsilon

    \displaystyle\frac{[ln\displaystyle\frac{\varepsilon+i\overline{\Gamma}/2]}{\varepsilon -i\overline{\Gamma}/2]}]}
    {[\varepsilon^{2}+\overline{\Gamma}^{2}/4]}=\\
    =\displaystyle\frac{4\pi^{2}\hbar^{2}(Trace \widetilde{P}(\varepsilon_{c},E_{0})\widetilde{P}(\varepsilon'_{c},E_{0}))}
    {\overline{\overline{D}^{2}}},
 \end{array}
\label{eq.59}
\end{equation}
where $\widetilde{P}(\varepsilon_{c},E),\widetilde{P}(\varepsilon'_{c},E),\widetilde{P}(\varepsilon_{c},E_{0})$   and
$\widetilde{P}(\varepsilon'_{c},E_{0})$ are the values of $\widetilde{P_{m}}$ and $\widetilde{P_{m'}}$ at points $\varepsilon_{c}, \varepsilon'_{c}$ and $E_{0}$ defined by the integral theorem around the average value, namely by the relations
\begin{equation}
    \int\limits_{-\infty}^{\varepsilon}d\varepsilon'
    \displaystyle\frac{\widetilde{P}(\varepsilon',E)}{\varepsilon'^{2}+\overline{\Gamma}/4}=
    \widetilde{P}(\varepsilon'_{c},E)\int\limits_{-\infty}^{+\infty}d\varepsilon'
    \displaystyle\frac{1}{\varepsilon'^{2}+\overline{\Gamma}/4},
\label{eq.60}
\end{equation}
\begin{equation}
\;
  \begin{array}{lcl}

     \int\limits_{-\infty}^{+\infty}d\varepsilon

    \displaystyle\frac{<Trace \widetilde{P}(\varepsilon,E)\widetilde{P}(\varepsilon'_{c},E)>}{\varepsilon^{2}+\overline{\Gamma}/4}
    ln\displaystyle\frac{\varepsilon+i\overline{\Gamma}/2}{\varepsilon-i\overline{\Gamma}/2}
    =\\
    =<Trace \widetilde{P}(\varepsilon,E)\widetilde{P}(\varepsilon'_{c},E)>
    \int\limits_{-\infty}^{+\infty}d\varepsilon
    \displaystyle\frac{ln\displaystyle\frac{\varepsilon+i\overline{\Gamma}/2}{\varepsilon-i\overline{\Gamma}/2}}
    {\varepsilon^{2}+\overline{\Gamma}/4}
 \end{array}
\label{eq.61}
\end{equation}
and
\begin{equation}
\;
  \begin{array}{lcl}

     <Trace \widetilde{P}(\varepsilon,E)\widetilde{P}(\varepsilon'_{c},E)>=

    \displaystyle\frac{\int|g(E)|^{2}Trace \widetilde{P}(\varepsilon,E)\widetilde{P}(\varepsilon'_{c},E)dE}
    {\int|g(E)|^{2}dE}

    =\\
    =Trace \widetilde{P}(\varepsilon,E)\widetilde{P}(\varepsilon'_{c},E)

 \end{array}
\label{eq.62}
\end{equation}
respectively.

     In order to obtain relations (58) and (59), we also used the residues Cauchy theorem for integrals
\[
  \oint\limits_{C_{0}}\displaystyle\frac{f(\varepsilon)}{\varepsilon-i\overline{\Gamma}/2}
   \ \ \ \ \ and \ \ \ \ \ \oint\limits_{C_{0}}\displaystyle\frac{f(\varepsilon)ln(\varepsilon \mp i\overline{\Gamma}/2)}
   {\varepsilon \pm i\overline{\Gamma}/2}d\varepsilon
\]
$f(\varepsilon)$ being an analytical function in the upper or lower half-plane of the complex values of $\varepsilon$  which vanishes over the half-circles $\Gamma_{\pm}$ with the infinitely large radius ; the contours $C_{0}, C_{1}$ and $C_{+}$ are represented in Fig.3 (the contours $\gamma_{-}$ and $\gamma_{+}$ on the edges of the cuts near the branch points of $ln(\varepsilon \pm i\overline{\Gamma}/2)$ are outside the integration contours). If one uses an analytical functions $D(E_{m})$ in the continuum approach instead of the mean spacing $\overline{D}$, the results do not change when one uses the approximation $D(E_{m}\pm i\Gamma_{m}/2)= \overline{D}$.

\begin{figure}[htb]
\centerline{\includegraphics[height=70mm]{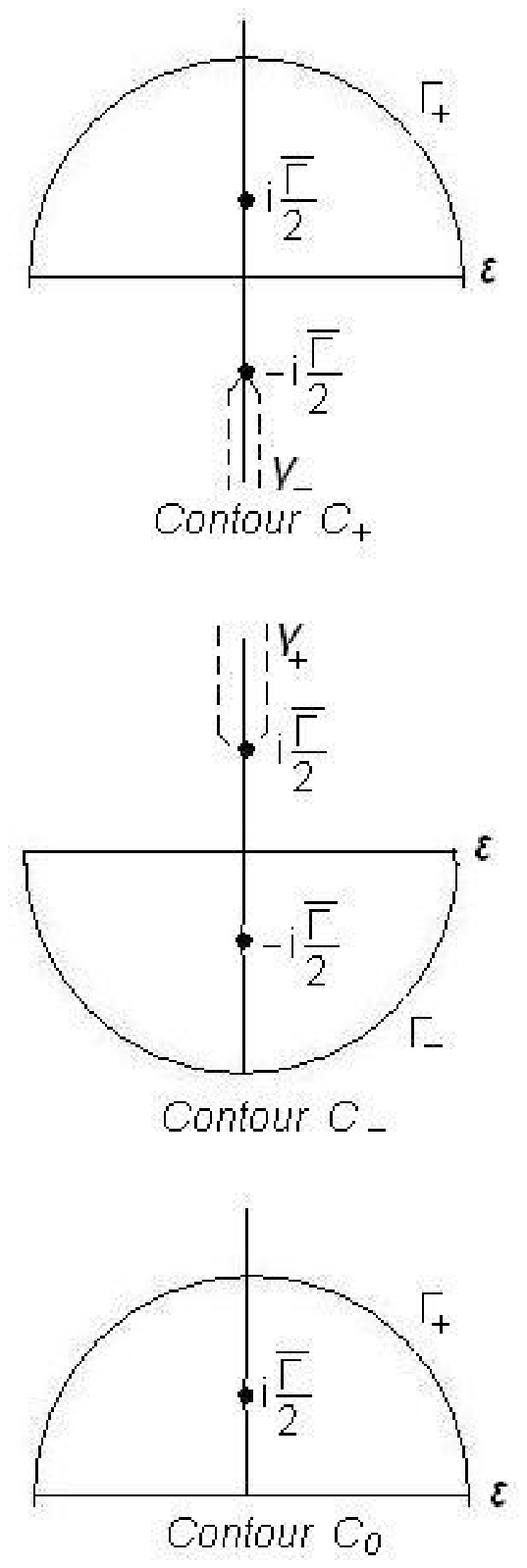}}
\caption{\small
Figure 3...
\label{fig.example.3}}
\end{figure}
Of course, the continuum approximation is justified only for averaging over $\Delta E >>\rho^{-1}$.

For statistically equivalent channels with $N >>1$ ($N\rightarrow\propto $ ) in the random-phase approximation (for $A_{mm',\nu}$), we shall make the following approximate evaluation
\begin{equation}
    Trace\widetilde{P}_{m}\cdot \widetilde{P}_{m'}\longrightarrow Trace\widetilde{P}(\varepsilon,E_{0})\cdot \widetilde{P}(\varepsilon'_{c},E_{0})\approx\sum\limits_{\nu}\displaystyle\frac{1}{N^{2}}=
    \displaystyle\frac{1}{N}.
\label{eq.63}
\end{equation}
Substituting the expressions (G.58) and (G.59) in (G.57), we finally obtain
\begin{equation}
    Re<Trace\widehat{R}_{res}>\cong\displaystyle\frac{4\pi\hbar^{2}\overline{\rho}}{\overline{\Gamma}}+
    4\pi^{2}\hbar^{2}\overline{\rho}^{2}\displaystyle\frac{M}{N},
\label{eq.64}
\end{equation}
with
\[
 \displaystyle\frac{M}{N}\equiv Trace\widetilde{P}(\varepsilon,E_{0})\widetilde{P}(\varepsilon'_{c},E_{0})
\]
which is equal to $1/N$  in the random-phase approximation for the statistically equivalent channels with $N >>1$ ($N\rightarrow\propto$).

Then, using again the $JS\Pi$ indexes, we obtain the following form of relation (64):
\begin{equation}
    Re<Trace\widehat{R}_{res}^{JS\Pi}> = 4\pi\hbar^{2}\rho^{JS\Pi}/\Gamma^{JS\Pi}+
    4\pi^{2}\hbar^{2}(\rho^{JS\Pi})^{2}M^{JS\Pi}/N^{JS\Pi}
\label{eq.65}
\end{equation}

Finally, comparing (G.52) with (G.65), we can write
\begin{equation}
\;
  \begin{array}{lcl}
     \sum\limits_{ik}[D\tau_{ik}^{JS\Pi}(E)+<\tau_{ik}^{JS\Pi}(E)>^{2}]/<|T_{ik}^{JS\Pi}(E)|^{2}>=4\pi\hbar^{2}\rho^{JS\Pi}/\Gamma^{JS\Pi}
     +\\
    + 4\pi^{2}\hbar^{2}(\rho^{JS\Pi})^{2}M^{JS\Pi}/N^{JS\Pi},
 \end{array}
\label{eq.66}
\end{equation}

In general, the quantity $M^{JS\Pi}/N^{JS\Pi}$ fluctuates between 0 and 1 (see relation (56) and can be used as an empirical parameter describing channel-channel correlations between resonances (more precisely, between resonant projection matrices $\widehat{P}_{m}$).

Now, for the study the compound-nucleus processes, we shall use the relations (omitting indexes $JS\Pi$)
\begin{equation}
    \displaystyle\frac{d|T_{k}|}{dE}=\displaystyle\frac{|T_{ik}^{c}|}{T_{ik}}\displaystyle\frac{d|T_{ik}^{c}|}{dE}+
    \displaystyle\frac{2}{|T_{ik}|}[Re<T_{ik}>\displaystyle\frac{d Re T_{ik}^{c}}{dE}+ Im<T_{ik}>
    \displaystyle\frac{d Im T_{ik}^{c}}{dE}]
\label{eq.67}
\end{equation}
and
\begin{equation}
    \displaystyle\frac{d\ arg\ T_{ik}}{dE}=\displaystyle\frac{|T_{ik}^{c}|^{2}}{|T_{ik}|^{2}}\displaystyle\frac{d\ arg\ T_{ik}^{c}}{dE}+\displaystyle\frac{\displaystyle\frac{d Im T_{ik}^{c}}{dE}Re<T_{ik}>-
    \displaystyle\frac{d Re T_{ik}^{c}}{dE}Im<T_{ik}>}{|T_{ik}|^{2}}
\label{eq.68}
\end{equation}
From relations (67) and (G.68), it follows that
\begin{equation}
    (\displaystyle\frac{d|T_{ik}|}{dE})^{2}+|T_{ik}|^{2}(\displaystyle\frac{d\ arg\ T_{ik}}{dE})^{2}=[(\displaystyle\frac{d|T_{ik}^{c}|}{dE})^{2}+|T_{ik}^{c}|^{2}(\displaystyle\frac{d\ arg\ T_{ik}}{dE})^{2}].
\label{eq.69}
\end{equation}
Hence, we can easily transform the sum rule (66), using (46), (47), (67), (68) and (69), into the form
\begin{equation}
\;
  \begin{array}{lcl}
     \sum\limits_{ik}[D\tau_{ik}^{c(JS\Pi)}(E)+<\tau_{ik}^{c(JS\Pi)}(E)>^{2}]/<|T_{ik}^{c(JS\Pi)}(E)|^{2}>=4\pi\hbar^{2}\rho^{JS\Pi}/\Gamma^{JS\Pi}
     +\\
    + 4\pi^{2}\hbar^{2}(\rho^{JS\Pi})^{2}M^{JS\Pi}/N^{JS\Pi}.
 \end{array}
\label{eq.70}
\end{equation}
In the approximation of the statistically equivalent channels for which the relation
\begin{equation}
    \sum\limits_{k}<|T_{ik}^{c(JS\Pi)}(E)|^{2}>\ \cong\ 1-exp(-2\pi\rho^{JS\Pi}\Gamma^{JS\Pi}/N^{JS\Pi})
\label{eq.71}
\end{equation}
is valid and all $<\tau_{ik}^{c(JS\Pi)}(E) >$ are equal to
\begin{equation}
    <\tau_{ik}^{c(JS\Pi)}(E) >=\displaystyle\frac{2\pi\hbar\rho^{JS\Pi}}{N^{JS\Pi}[1-exp(-2\pi\rho^{JS\Pi}\Gamma^{JS\Pi}/N^{JS\Pi})]},
\label{eq.72}
\end{equation}
all $(<\tau_{ik}^{c(JS\Pi)}(E)>)^{2}$ are also equal to each other.
For the case of weakly overlapping resonances, when $\Gamma^{JS\Pi}<<N^{JS\Pi}/2\pi\rho^{JS\Pi}$,
\[
 <\tau^{c(JS\Pi)}>=\hbar/\Gamma^{JS\Pi}\ \ \ and\ \ \ D\tau^{c(JS\Pi)}=(\hbar/\Gamma^{JS\Pi})^{2}=
 (<\tau^{c(JS\Pi)}>)^{2},
\]
i.e., in total agreement with [13,15] the compound-nucleus decay law is, on average, exponential.

For the case of strongly overlapping resonances, when  $\Gamma^{JS\Pi}>>N^{JS\Pi}/2\pi\rho^{JS\Pi} $ ,
\begin{equation}
    <\tau^{c(JS\Pi)}>=2\pi\rho^{JS\Pi}/N^{JS\Pi}\ \ \ and\ \ \ D\tau^{c(JS\Pi)}\approx
    (2\pi\hbar\rho^{JS\Pi}/N^{JS\Pi})^{2}\ \ \ll\ \ (\tau^{c(JS\Pi)})^{2}
\label{eq.72}
\end{equation}
hence also in agreement with [14,16,17] the compound-nucleus decay law is strongly non-exponential, i.e. is a time resonance (explosion).

\section{COMPARISON WITH SOME EXPERIMENTAL DATA.
\label{sec.4}}

For analysis of observed experimental spectra of a single final fragment one has to sum (or average) the expressions like (33) and (35) over the subsets of final states (with different quantum numbers $JS\Pi$, orbital quantum numbers \emph{L} etc) and channels, sometimes coherently (see[25]) and usually incoherently. It does not matter and does not change the final expressions if we make the simplifying assumption that all the involved amplitudes have the same \emph{E}-dependence for the both $|f_{\alpha\beta}|$ and  $\arg f_{\alpha\beta}$ . In particular, for inclusive energy spectra of the \emph{k}-th final fragment we shall use the following expression:
\begin{equation}
  \;
  \begin{array}{lcl}
     \sigma_{inc,k}(E_{k})=|\sum\limits_{n=1}^{2}C_{n}\exp[(it_{n}-\tau_{n}/2)E_{k}/\hbar]|^{2}=
     \\
    =\sum\limits_{n=1}^{2}|C_{n}|^{2}\exp(-E_{k}\tau_{n}/\hbar)+
    \\
    +  2Re  C_{1} C_{2}\exp\{[i(t_{2}-t_{1})-(\tau_{2}+\tau_{1})/2]E_{k}/\hbar\}
  \end{array}
\label{eq.54}
\end{equation}

In Fig. 4-7 some calculated inclusive energy spectra $\sigma_{inc,k}(E_{k})$, in arbitrary units and in semi-logarithmic scale, are presented in comparison with the experimental data from [4,6,8,9,11,12]. At the pictures the experimental data are the points, and the theoretical data are the curves.

\begin{figure}[htbp]
\centerline{
\includegraphics[width=165mm]{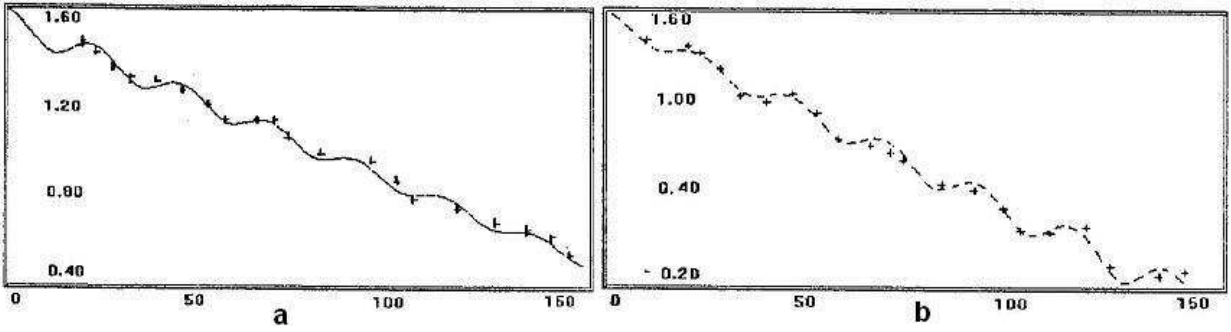}}
\caption{\small
 Inclusive process $p + C \rightarrow Be^{7}$ (2.1 \emph{GeV} protons); the experimental points are taken from [11]. a) C1= 0.04, C2= 0.36 ($\theta = 90^{°}$); b) C1= 0.35, C2= 0.05 ($\theta = 160^{°}$)
\label{fig.example.3}}
\end{figure}

\begin{figure}[htbp]
\centerline{
\includegraphics[width=145mm]{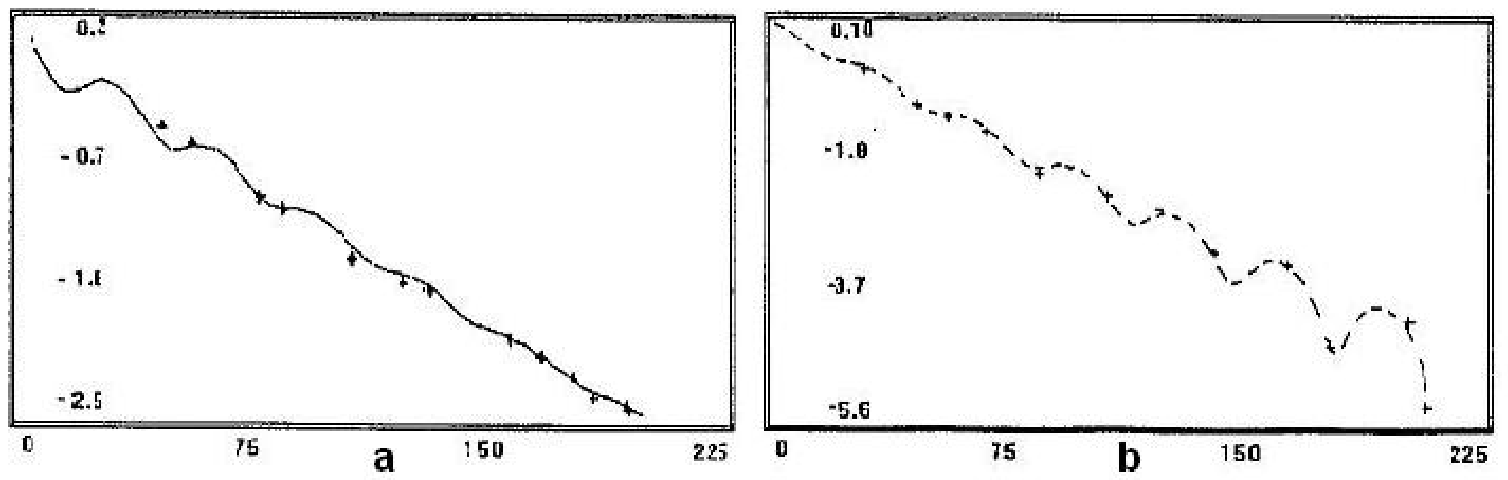}}
\caption{\small
 Inclusive process $Ne^{20} + Al \rightarrow p$ (393 \emph{MeV} nucleons); the experimental points are taken from [6]. a) C1= 0.2, C2= 5.8 ($\theta = 90^{°}$); b) C1= 5.8, C2= 0.2 ($\theta = 150^{°}$)
\label{fig.example.4}}
\end{figure}

\begin{figure}[htbp]
\centerline{
\includegraphics[width=150mm]{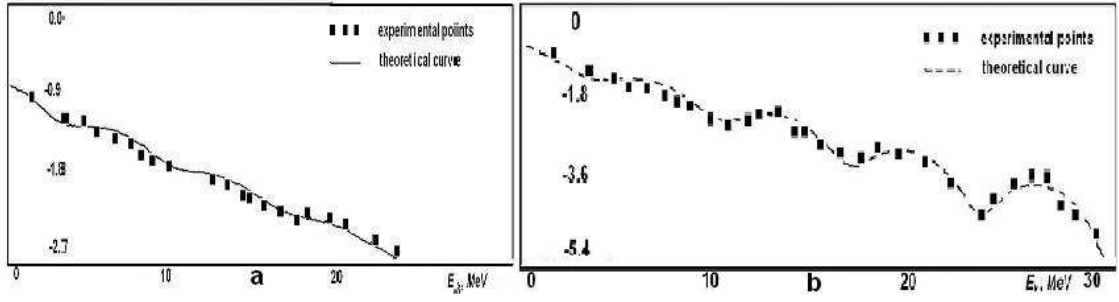}}
\caption{\small
 Inclusive process $He^{4} + t \rightarrow p$ (720 \emph{MeV} nucleons); the experimental points are taken from [8]. a) C1= 0.18, C2= 1.02 ($\theta = 60^{°}$); b) C1= 1.13, C2= 0.07 ($\theta = 90^{°}$)
\label{fig.example.5}}
\end{figure}

\begin{figure}[htbp]
\centerline{
\includegraphics[width=150mm]{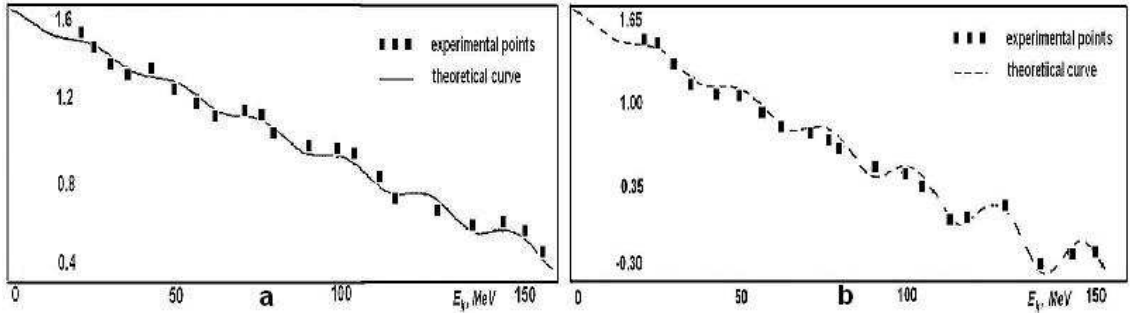}}
\caption{\small
 Inclusive process $Ne^{20} + Ul \rightarrow p$ (1045 \emph{MeV} nucleons); the experimental points are taken from [8]. a) C1= 0.35, C2= 5.65 ($\theta = 90^{°}$); b) C1= 5.65, C2= 0.35 ($\theta = 150^{°}$)
\label{fig.example.6}}
\end{figure}

\begin{figure}[htbp]
\centerline{
\includegraphics[width=150mm]{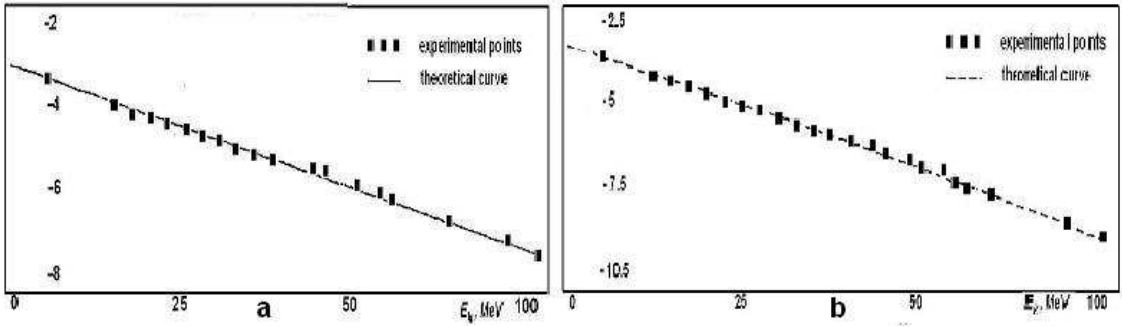}}
\caption{\small
 Inclusive process $Ar^{40} + V^{51} \rightarrow p$ (41 \emph{MeV} nucleons); the experimental points are taken from [9]. a) C1= 0.002, C2= 0.03 ($\theta = 97^{°}$); b) C1= 0.03, C2= 0.022 ($\theta = 129^{°}$)
\label{fig.example.7}}
\end{figure}

In Fig.4-7, $\theta$ is the detected angle of the \emph{k}-fragment emission.

The values of $\tau_{1}, \tau_{2}$ and $t_{2} - t_{1}$, which were found from fitting the theoretical curves with the experimental data, are presented in Table 1.


\renewcommand{\tablename}{Table}
\begin{table}
\begin{center}
\begin{tabular}{|c|c|c|c|c|} \hline
  Reaction &
  Projectile energy, \emph{Gev/nucl} &
  $\tau_{1},  10^{-23} sec$ &
$\tau_{2},  10^{-23} sec$ &
$t_{2}-t_{1},  10^{-22} sec$ \\ \hline
  $p + C \rightarrow Be^{7}$&
$2.1 $ &
$10.45 $ &
$17.0 $ &
$5.95$ \\ \hline
  $Ne^{20} + Al \rightarrow p$&
$0.393 $ &
$0.1 $ &
$0.99 $ &
$1.7$ \\ \hline
  $He^{4} + Ta \rightarrow t$&
$0.72 $ &
$1.72 $ &
$3.15 $ &
$1.22$ \\ \hline
  $Ne^{20} + U \rightarrow p$&
$1.045 $ &
$0.92 $ &
$1.7 $ &
$1.72$ \\ \hline
  $Ar^{20} + V \rightarrow p$&
$0.041 $ &
$7.5 $ &
$9.0 $ &
$0.20$ \\ \hline
  $Xe^{132} + Au \rightarrow p$&
$0.044 $ &
$6.0 $ &
$7.0 $ &
$1.0$ \\ \hline
  $Ne^{20} + U \rightarrow p$&
$0.4 $ &
$1.7 $ &
$2.2 $ &
$0.10$ \\ \hline
  $Ne^{20} + U \rightarrow d$&
$0.25 $ &
$4.2 $ &
$7.2 $ &
$0.10$ \\ \hline
\end{tabular}
\end{center}
\caption{. The parameters of time resonances for some inclusive spectra.
\label{table.1}}
\end{table}

If in [8] the attempt was made to seek the same one-two time resonances for a wide region of energies e and total numbers of nucleons involved in collisions, we had searched the same one-two time resonances only for the same (or almost the same) energy excitations of the same (or almost the same) compound systems (nuclei or clots). And only the contributions \emph{C1} and \emph{C2} were varied for various emission angles and fragment masses.

Since the slope of energy spectra noticeably increases with increasing angle, it signifies that the increasing contribution of compound-nucleus states with larger values of $t_{n}$ and $\tau_{n}$ is connected with the formation of a heavier compound nucleons with a less velocity (in the laboratory system). This agrees with the observed in [1,3,6] phenomena of more distinct oscillations for intermediate emission angles.

May be for the lightest compound system (p+C), presented here, the superposition of a direct process (i.e. \emph{n = 0} instead of \emph{n = 1}) and a time resonance (\emph{n = 2}) is taking  place because the difference $t_{2} - t_{1(0)}$ is larger than usual.

\section{CONCLUDING REMARKS
\label{sec.conclusion}}

Here was developed the phenomenological time-evolution approach, which is based on general properties of time-energy transformations and general results of joint time and statistical energy-resonance analysis of nuclear reactions, obtained earlier in [14-18], even without taking the formal considerations, developed in [8,13] as, in a certain degree, a continuation of that analysis - and a further development of the time-operator formalism seems to be very useful for elaborating the microscopic models of time evolution of high-energy collision processes with utilizing concrete properties of time operators, canonically conjugated to compound-nucleus Hamiltonians. At the same time, relative to our opinion, presented here time-resonance (explosion) phenomenological description can also be combined with any semi-microscopic model, in particular, elaborated on the base of the fireball or intra-nuclear cascade models.

In section 3 the certain situations at the range of strongly overlapped compound-resonances were revealed (in particular, conditions $\pi \Gamma\rho \gg 1$ and (49), then $\pi \rho\Gamma /N \gg 1$ etc) when one can observe one or several time resonances (explosions) in the cross sections \emph{de facto}. So, time resonances (explosions) can appear only on the base of strongly overlapped standard energy compound-resonances as a particular phenomenon. It is worth to mark that the only clear experimental indication for appearing of a time resonance (explosion) is the exponential decreasing of inclusive energy spectra (with some oscillations at the case of several coherent time resonances [for the same involved amplitudes]).

At present, because of somewhat unclear (rigidly in practice) validity of the conditions (49) and even (52)-(53) it is now difficult to conclude unequivocally, basing on selected good results of fitting in Figs 3-4, if the time resonances are really existing or not. However, the time-resonance (explosion) phenomenon is a rather sane hypothesis. Further, we need more accurate and precise experimental data for a certainty in the presence of real oscillations but not of random dispersions and errors of measurement results.

As to the dependences of the cross sections and inclusive energy spectra on the detected angle of the final fragment emission, in general, one can expect the same picture as for typical compound-nucleus  processes, with some particular exceptions like the following. It seems that a certain experimental indication for the existence of time resonances can be seen in the phenomena of the steeper energy decreasing in spectra for the larger emission angles $\theta$ and of the more distinct oscillations in spectra for the intermediate (between the minimal and maximal experimental values) emission angles $\theta$ in the laboratory system which are observed in [1-6]. Nevertheless, one has to consider that for small angles the last oscillation peak in spectra (for the largest energies \emph{E}) can be connected with the pure kinematical effect of direct or peripherical processes in the laboratory system. But at least one very important conclusion we can already state - about a new kind of analysis: joint time-evolution and statistical energy-resonance approach for the study of dependencies of energy-resonance densities, widths and numbers of open channels from $JS\Pi L$ and \emph{E} for unresolved (and consequently unobservable) compound-nucleus energy resonances, and also for the study of possible direct-process contributions. As to dependencies from $JS\Pi L$, one can hope that the investigation of energy spectra for various emission angles will be useful.

\renewcommand{\refname}{Reference:}

\end{document}